\documentclass{chjaa}                  

\usepackage{graphicx,times}             
\input{epsf.sty}                        
\input{psfig.sty}                       

\begin{document}

   \title{Does Amati Relation Depend on Luminosity of GRB's Host Galaxies?
}

   \volnopage{Vol.0 (200x) No.0, 000--000}      
   \setcounter{page}{1}           

   \author{Jing Wang, Jing-song Deng and Yu-lei Qiu
      \inst{1}
      }

   \institute{National Astronomical Observatories, Chinese Academy of Sciences,
             Beijing 100012, China\\
             \email{wj@bao.ac.cn}
          }

   \date{Received~~2007 month day; accepted~~2007~~month day}

   \abstract{
   In order to test systematic of the Amati relation,  
   the 24 long-duration GRBs with firmly determined $E_{\gamma,\mathrm{iso}}$ and $E_{\mathrm p}$
   are separated into two sub-groups according to
   B-band luminosity of their host galaxies. The Amati relations in the two subgroups 
   are found to be in agreement with each other within uncertainties.  
   Taking into account of the well established luminosity - metallicity
   relation of galaxies, no strong evolution of the Amati relation with GRB's 
   environment metallicity is implied in this study.
   }

   \authorrunning{J. Wang et al}            
   \titlerunning{Does Amati Relation Depend on Host's Luminosity}  

   \maketitle

%
%
\section{Introduction}           
\label{sect:intro}

Several relations between observable properties have been found for 
Gamma Ray Bursts (GRBs) observed
in past few years (see Schaefer 2007 for a review). Among these relations, 
an important one is the Amati relation that is a correlation between 
the total isotropic-equivalent radiated energy in $\gamma$-ray ($E_{\gamma,\mathrm{iso}}$) of 
long-duration GRBs (LGRBs) and their peak energy ($E_{\mathrm{p}}$) of 
integrated $\nu F_\nu$ spectrum in the rest frame (Amati et al. 2002). This 
correlation is further confirmed and extended by subsequent observations
(e.g., Amati 2003, 2006a,b; Sakamoto et al. 2004; Lamb et al. 2004). 
A similar relation between $L_{\mathrm{iso}}$ and $E_{\mathrm p}$ is found to be not only 
hold among different LGRBs but also hold among individual pulses of single LGRB (Liang et al. 2004).   
Although the correlation is highly significant, the dispersion of the correlation 
is expected to be not only caused by statistic fluctuation (e.g., Amati 2006a). 

At present, LGRBs are generally believed to originate 
from the death of young massive stars (e.g., see Woosley \& Bloom (2006) for a recent 
review). The popular collapsar model favors progenitors with low metallicity to preserve angular 
momentum when the collapse occurs (e.g., Woosley 1993; MacFadyen \& Woosley 1999).
A low metallicity environment has been 
indeed reported by the studies of host galaxies of both nearby and 
cosmologically distant LGRBs (e.g., Sollerman et al. 2005; Stanek et al. 2006; Fynbo et al. 2006).   
Because metal abundance plays an important role in the collapsar model, the evolution
of the statistical properties of LGRB is therefore expected. 

Li (2007) recently examined the cosmological evolution of the Amati relation by dividing 48 LGRBs with
reported $E_{\mathrm{iso}}$ and $E_{\mathrm p}$
into four redshift bins. The Amati relation is found to vary with redshift with only $\sim$4\% 
of chance that the variation is caused by selection effect. 
Although it is generally believed that metallicity statistically evolves strongly with redshift, 
a number of extremely metal-poor galaxies,
have been identified in the local Universe (e.g., Kewley et al., 2007; Izotov et al., 2006 
and references therein). Since the metallicity is hard to be determined for a large sample of LGRBs at 
current, the luminosity (or stellar mass) of
host galaxy could be used as a physically meaningful indicator of metallicity 
taking into account of the well established 
luminosity (or mass)-metallicity relationship ($L-Z$ relation, e.g., Tremonti et al., 2004; 
Savaglio et al. 2005). 
In this paper, we examine the variation of the Amati relation on 
luminosity of LGRB's host galaxy.


\section{Variation of Amati Relation on Luminosity of Host Galaxy}
\label{sect:Obs}
We compile a sample of LGRBs to examine whether the Amati relation varies with luminosity of host 
galaxy. The B-band luminosities of host galaxies are adopted from published literature, and is 
transformed to absolute B-band magnitude by adopting $M_B^{\star}=-21$ mag. In order to 
avoid the selection effect, only the LGRBs with $0.2<z<2$ are considered, which leads the 
four nearby bursts, i.e., GRB\,980425, GRB\,030329, GRB\,031203 and GRB\,060218, 
are excluded from our sample. GRB\,980425
is a sub-energetic LGRB and dose not found to satisfy the Amati relation. GRB\,031203 has 
poorly determined $E_{\mathrm p}$. The sample finally 
contains 24 LGRBs listed in Amati (2006). Table 1 lists the properties for each LGRB,
including the redshift, rest-frame isotropic energy $E_{\gamma,\mathrm{iso}}$ defined in 1-1000keV band,
peak energy $E_{\mathrm{p}}$ in rest-frame, and $k$-corrected absolute B-band magnitude $M_B$ of its host galaxy.
The $\Lambda$CDM cosmology with $\Omega_{\mathrm m}=0.3$, $\Omega_\Lambda=0.7$ and 
$h_0=0.7$ is adopted throughout the paper.

The redshift is plotted against $M_B$ for our sample in Figure 1 (left-bottom panel). The diagram indicates 
that there is no clear trend of $M_B$ on redshift in the range from $z=0.2$ to $z=2$. 
The Figure 1 upper panel shows the distribution of $M_B$ of the 24 LGRBs. 
The $M_B$ spans a range of $-16$ - $-22$ mag.  In order to examine variation 
of the Amati relation on luminosity of LGRB's host, we separate the sample into two subgroups with 
identical contents, 
i.e., Group L and H, according to the 
luminosity of LGRB's host galaxy. The LGRBs with $M_B>-19.7$ belong to Group L, and 
these with $M_B\leq-19.7$ Group H (see vertical dashed line in Figure 1). 
Finally, each group contains 
 12 LGRBs. The bottom-right panel of Figure 1 shows the distributions of redshift for both subgroups 
(solid line for Group H and dashed line for Group L). In both subgroups, a majority of 
LGRBs are distributed in a narrow range from $z=0.5$ to 1. 
A logrank test indicates that the redshift distributions of the two subgroups are 
drawn from the same parent population at a probability $\sim70$\%.

Using 41 LGRBs with firmly determined $E_{\gamma,\mathrm{iso}}$ and $E_{\mathrm{p}}$, Amati (2006)
obtained an updated relationship 
\begin{equation}
\log E_{\gamma,\mathrm{iso}}=a+b\log E_{\mathrm{p}}
\end{equation}
where $a=-3.35$ and $b=1.75$ using the least squares fitting method; 
and $a=-4.04$ and $b=2.04$ using the 
maximum likelihood method. A least squares fit to our 24 LGRBs as a single sample with Eq. (1) leads to
$a=-3.25\pm0.40$ and $b=1.69\pm0.16$ with $\chi_r^2=1.60$. The $\chi_r^2$ is the reduced $\chi^2$, which is
defined as the $\chi^2$ of fit divided by the degree of freedom. The fitting is shown in Panel A in 
Figure 2. The two dashed lines in the Panel A mark the 1$\sigma$ deviation of the best fit.  
This result is in good agreement with that obtained by Amati (2006) and Li (2007), which indicates
that no additional bias is obviously introduced in the sample used in this paper by our sample selection. 

Least squares fittings are also carried out for both Group H and L. 
The fittings are shown in Panel B and C in 
Figure 2 for Group H and L, respectively. In Group H, the best-fit Amati relation has parameters $a=-2.97\pm0.73$ and
$b=1.61\pm0.30$ (with $\chi_r^2=1.96$), which is similar to $a=-3.47\pm0.44$ and $b=1.74\pm0.17$ 
(with $\chi_r^2=0.94$) obtained in Group L within uncertainties.
In addition to the best fit of the Amati relation, the 
dispersion around the best fit also provides important information. Although the dispersion of Group H
is slightly larger than that of the whole sample, the dispersion of Group L is found to be
significantly suppressed (see also the $\chi^2_r$ for each group). Figure 3 plots the distributions
of deviation from the best fit in $\log E_{\gamma,\mathrm{iso}}$. The distribution for the whole sample,
Group H and Group L is shown in Panel A, B and C, respectively. As shown in the Panel C, 
the distribution of Group L is quite uniform with an obvious cut-off at $\sim$0.6. In contrast, 
Group H shows a relatively wider distribution with a clear peak at $\sim$0.5.
The difference in distribution of dispersion confirms the separation of the 24 LGRBs into the two subgroups, 
although the origin of the difference is out of the scope of this paper.

\section{Discussion and Summary}

The distributions of $\log E_{\gamma,\mathrm{iso}}$ for both subgroups are shown in 
right-bottom panel in Figure 2. The symbols are the same as that in the right panel of Figure 1.
Both subgroups have similar dynamical range of $\log E_{\gamma,\mathrm{iso}}$, 
a sub-luminous LGRB (GRB\,020903) is, however, only found in Group L. 
The Amati relation is also fitted through Eq. (1) after excluding the sub-luminous GRB\,020903 from
Group L. We obtain a relation with $a=-3.48\pm0.96$ and $b=1.74\pm0.36$ ($\chi_r^2=1.33$),
which confirms the consistence of our fitting.

The $L-Z$ relation has been firmly established in the local Universe ($z<1$)
base upon various spectroscopic surveys (e.g., Tremonti et al. 2004; Savaglio et al. 2005; 
Liang et al. 2006). The $L-Z$ relation indicates that, in general, 
high metallicity is found in luminous galaxies, 
and low metallicity in faint galaxies. Tremonti et al. (2004) obtained a relationship
$12+\log(\mathrm{O/H})=-0.185M_B+5.238$ with a typical scatter of $\sigma_{12+\log(\mathrm{O/H})}=0.16$
from SDSS. The existent observations indicate 
that local LGRBs' host galaxies are not far from the $L-Z$ relation (e.g., Savaglio et al. 2006).
The median value of absolute magnitude is $M_{\mathrm{B}}=-20.35$ mag for Group H, and -18.55 mag for Group L.
According to the relationship derived by Tremonti et al. (2004), the difference of 
metallicity is inferred to be 0.33 dex which is two times larger than the dispersion of the $L-Z$ relation.
Recent observations revealed an evolution of the zero point of the $L-Z$ relation from 
local Universe to intermediate redshift $z=1$. Different evolution are, however, found by 
various authors. For instance, Liang et al. (2006) found an evolution of 0.3 dex since $z=0.65$, while 
a much more moderate evolution of 0.14 dex is reported by Kobulnicky \& Kewley (2004).
In current study, the cosmological evolution is not a key issue 
because a majority of LGRBs in \it both \rm two sub-groups
are uniformly distributed in a relatively narrow dynamical range of redshift 
(from $z=0.5$ to $z=1$, see Figure 1).
According to these existent observations, the consistence of the Amati relations for different 
luminosity of host galaxy implies that the Amati relation has no strong evolution 
with metallicity of LGRB's environment.

In the generally accepted fireball model, the Amati relation could be explained by the standard internal 
shock scenario, $E_{\mathrm p}\propto \Gamma^{-2}L^{1/2}t^{-1}_{\mathrm{var}}$, where $\Gamma$ is the fireball 
bulk Lorentz factor, $L$ is the GRB luminosity and $t_{\mathrm{var}}$ is the typical variability time scale 
(e.g., Zhang \& M\'{e}sz\'{a}ros 2002; Rees \& M\'{e}sz\'{a}ros 2005; Ryde 2005). 
The agreement of the Amati relations in the two subgroups would consequently require that both $\Gamma$ 
and $t_{\mathrm{var}}$ are approximately independent on environment metallicity. An alternative
explanation of the Amati relation is the thermal radiation from photosphere of GRB 
(e.g, Rees \& M\'{e}sz\'{a}ros 2005; Thompson 2006; Thompson et al. 2007). In such a context, one expects 
$E_{\mathrm{p}}\propto R_0^{-1/2}\Delta t_j^{-1/4}E_{\gamma,\mathrm{iso}}^{1/2}$, where $R_0$ is the 
radius of complete thermalization. Such radius is reasonably assumed to be comparable or less than 
the radius of core of the progenitor. Because of the weak dependence on the duration of prompt emission,
the slope of the Amati relation primarily depends on the radius. 
Our test therefore implies a similar core radius of progenitor in both subgroups.

The result obtained 
in this paper differs from that in Li (2007). Li (2007) found 
the variation of the Amati relation by separating the whole sample into four groups according 
to redshift. However, various selection bias should be carefully considered in such study. 
In stead of the $E_{\mathrm p}$-$E_{\gamma,\mathrm{iso}}$
relation, a much lower dispersion is found in the $E_{\mathrm p}$-$E_{\gamma}$ relation by 
correcting for collimation angles of jet (Ghirlanda et al. 2004; 2007). 
The different slope between the $E_{\mathrm p}$-$E_{\gamma,\mathrm{iso}}$ and 
$E_{\mathrm p}$-$E_{\gamma}$ relation leads to a hypothesis that powerful 
bursts intrinsically have smaller opening angles (Ghirlanda et al. 2005). Assuming the 
$E_{\mathrm p}$-$E_{\gamma}$ relation is intrinsic for all LGRBs, the dependence of 
opening angle on burst energy could be a possible explanation of the 
decrease of slope of the Amati relation with redshift, because sub-luminous 
bursts have been only detected in local Universe by now.

As done in Ghirlanda et al. (2005), assuming the $E_{\mathrm p}$-$E_{\gamma}$ relation is 
intrinsic for all the LGRBs, it is possible to compare the properties of the burst environment 
in the two subgroups. So far, the uncertainties of burst environment have not been considered in
the previous studies on the $E_{\mathrm p}$-$E_{\gamma}$ relation. Moreover, the 
properties of the circumburst medium could provide some insight about the energy source of LGRB. 
The model of LGRB's afterglow lighcurves indicates that a homogeneous medium is more favored 
than a wind like $r^{-2}$ radial stratification (e.g, Panaitescu 2005; and recently 
summarized in Fryer et al. 2007). The observed homogeneous medium could be explained by either  
termination shock of wind (Wijers 2001) or bubbles with uniform density produced in intense 
starburst region (Chevalier et al. 2004).

In the homogeneous case, the density is (Sari 1999)
\begin{equation}
n=\frac{E_{\gamma,\mathrm{iso,52}}}{\eta}\bigg(\frac{\theta_{\mathrm{j}}}{0.161}\bigg)^8
\bigg(\frac{t_{\mathrm j}}{1+z}\bigg)^{-3}\ \mathrm{cm^{-3}}
\end{equation}
where $\theta_{\mathrm j}$ is the opening angle of jet, $\eta$ is the radiation efficiency which 
is usually assumed to be the same for all bursts, $\eta=0.2$ (Frail et al. 2001) and 
$t_{\mathrm j}$ is the break time in units of day of the afterglow light curve. 
The estimated density is
shown in Column (7) in Table 1 for the 11 LGRBs with firm estimates of jet break time
(as shown in Column 6 in Table 1). 
For all the 11 LGRBs, the inferred values of density are consistent with the observations. 
The measured density roughly extends from 1 to 10 $\mathrm{cm^{-3}}$ (e.g., Frail et al. 2001; 
Panaitescu \& Kumar 2002; Schaefer et al. 2003). 
Comparing the density between Group H and Group L, 
it is noted that the distribution of density in Group L is roughly concentrated 
around the value of $2\ \mathrm{cm^{-2}}$. 
On the contrary, the density spreads wider in Group H than in Group L. 

We note here that our treatment of binning the sample into two subgroups according to 
B-band luminosity of host galaxies is more or less simplified. 
The B-band luminosity is not a perfect indicator of stellar mass (or 
metallicity) because it suffers from extinction and traces mass of only 
massive stars. More accurate treatment of binning and direct measurement of metallicity 
are required to accurately test the evolution of the Amati relation.

In summary, we examine the systematics of the Amati relation by dividing the 24 LGRBs into 
two subgroups according to the absolute B-band magnitudes of 
their host galaxies. Obvious difference of Amati relations in the two subgroups are not found 
within uncertainties, 
although they seem to have different distributions of deviation from the best fit. 
Combining with the well established luminosity - metallicity relation, current study does not 
imply strong evolution of the Amati relation with LGRB's environment metallicity.

%
\begin{table}[]
  \caption[]{The sample of 24 LGRBs used in this work}
  \label{Tab:publ-works}
  \begin{center}\begin{tabular}{ccccccccc}
  \hline\hline\noalign{\smallskip}
GRB & z & $E_{\gamma,\mathrm{iso}}$ & $E_{\mathrm{p}}$ & $M_B$ & $t_{\mathrm j}$ & $n$ &  Ref\\
    &   &  ($\mathrm{erg\ s^{-1}}$) &    (keV)         &  (mag)  &  (day)  &  ($\mathrm{cm^{-3}}$) & \\
(1) & (2) & (3) & (4) & (5) & (6) & (7) & (8) \\
  \hline\noalign{\smallskip}
970228 & 0.695 & $1.86\pm0.14$ & $195\pm64$ & -17.85 &  \dotfill & \dotfill &  1,2\\
970508 & 0.835 & $0.71\pm0.15$ & $145\pm43$ & -17.85 &  \dotfill & \dotfill &  1,2\\
970828 & 0.98 & $34\pm4$ & $586\pm117$ & -18.85 &   2.2 & 2.04 &  1,2,7 \\
980613 & 1.096 & $0.68\pm0.11$ & $194\pm89$ & -19.85 &  \dotfill & \dotfill &1,2\\
980703 & 0.966 & $8.3\pm0.8$ & $503\pm64$ & -20.90 &   3.4 & 4.18 &  1,2,7\\
990123 & 1.60 & $266\pm43$ & $1724\pm466$ & -20.40 &   2.04 & 3.43 & 1,2,8\\
990506 & 1.30 & $109\pm11$ & $677\pm156$ & -19.75 & \dotfill & \dotfill & 1,2\\
990510 & 1.619 & $20\pm3$ & $423\pm42$ & -17.20 &  1.6 & 2.50 & 1,2,7\\
990705 & 0.842 & $21\pm3$ & $459\pm139$ & -21.65 & 1.0 & 1.01 & 1,2,7\\
990712 & 0.434 & $0.78\pm0.15$ & $93\pm15$ & -19.35 & 1.6 & 1.21 & 1,2,7\\
991208 & 0.706 & $25.9\pm2.1$ & $313\pm31$ & -18.30 & \dotfill & \dotfill & 1,2\\
991216 & 1.02 & $78\pm8$ & $648\pm134$ & -18.15 &  1.2 & 1.94 & 1,2,7\\
000210 & 0.846 & $17.3\pm1.9$ & $753\pm26$ & -19.50 &  \dotfill & \dotfill & 1,2\\
000418 & 1.12 & $10.6\pm2.0$ & $284\pm21$ & -19.90 & \dotfill & \dotfill & 1,2\\
000911 & 1.06 & $78\pm16$ & $1856\pm371$ & -18.80 & \dotfill & \dotfill & 1,2\\
010921 & 0.45 & $1.10\pm0.11$ & $129\pm26$ & -19.75 &  \dotfill & \dotfill & 1,2\\
011121 & 0.36 & $9.9\pm2.2$ & $793\pm533$ & -16.15 &  \dotfill & \dotfill & 1,2\\
020405 & 0.69 & $12.8\pm1.5$ & $612\pm122$ & -21.50 &  1.67 & 29.49 & 1,2,7\\
020813 & 1.25 & $76\pm19$ & $590\pm151$ & -19.30 &   0.43 & 2.04 & 1,2,7\\
020903 & 0.25 & $0.0028\pm0.0007$ & $3.37\pm1.79$ & -19.2 &  \dotfill & \dotfill& 1,3\\
030328 & 1.52 & $43.0\pm4.0$ & $328\pm55$ & -20.4 &  0.8 & 0.58 & 1,4,7\\
030528 & 0.782 & $2.0\pm0.7$ & $57\pm9$ & -21.4 &  \dotfill & \dotfill & 1,5\\
050223 & 0.5840 & $10\pm4.6$ & $109.6\pm60.6$ & -20.0 &  \dotfill & \dotfill &  1,6\\
050416 & 0.65 & $0.12\pm0.02$ & $25.1\pm4.2$ & -20.3 & 1.0 & 1.17 & 1,9\\

  \noalign{\smallskip}\hline
  \end{tabular}\end{center}
\note{References: 1. Amati 2006a; 2. Le Floc'h et al. 2003; 3. Hammer et al. 2006; 
4. Gorosabel, J., et al. 2005; 5. Rau et al. 2005; 6. Pellizza et al. 2006;
7. Ghirlanda et al. 2007; 8. Ghirlanda et al. 2004; 9. Soderberg et al. 2007}
\end{table}

%
\begin{figure}
   \vspace{2mm}
   \begin{center}
   \hspace{3mm}\psfig{figure=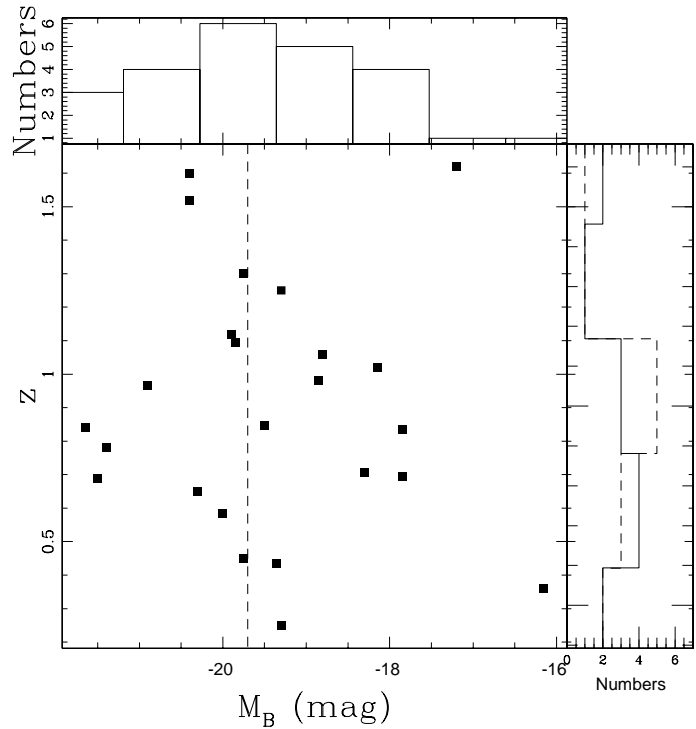,width=100mm,height=100mm,angle=0.0}
   \caption{\it Bottom-left panel: \rm absolute B-band magnitude of LGRB's 
   host galaxies plotted against redshift. The vertical dashed line 
   marks the point $M_B=-19.7$ which is used to divide the 24 LGRBs into 
   two subgroups (see text for details). \it Upper panel:\rm the distribution of 
   absolute B-band magnitude of LGRB's host galaxies. \it Bottom-right panel: \rm
   distributions of redshift for the two subgroups (solid line for Group H, and dashed 
   line for Group L).}
   \label{Fig:lightcurve-ADAri}
   \end{center}
\end{figure}

\begin{figure}
   \vspace{2mm}
   \begin{center}
   \hspace{3mm}\psfig{figure=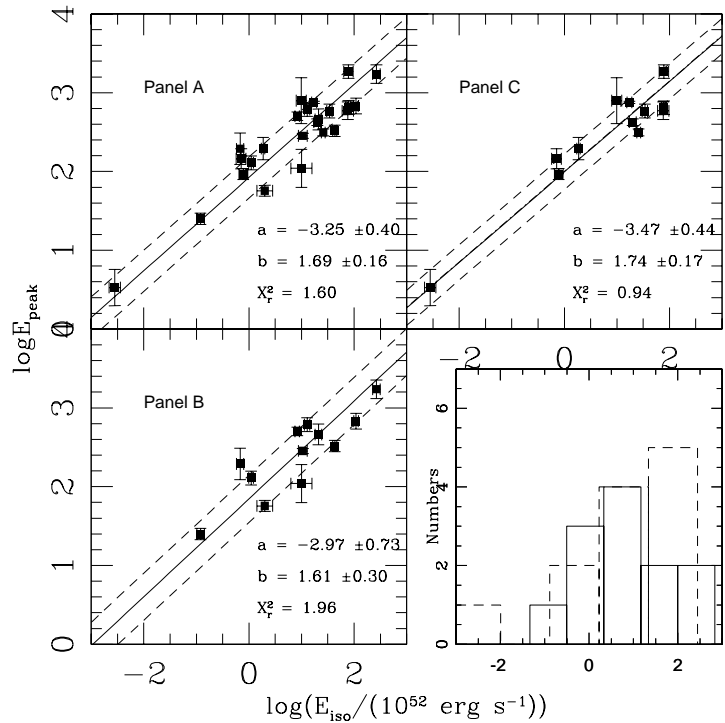,width=100mm,height=100mm,angle=0.0}
   \caption{Least squares fits (solid lines) to the 24 LGRBs as a single sample (\it{Panel A}\rm), to 
Group H (\it{Panel B}\rm) and Group L (\it{Panel C}\rm). The two dashed lines in each panel 
mark the 1$\sigma$ deviation from the best fit. \it Right-bottom Panel\rm: Distributions of 
$\log E_{\gamma,\mathrm{iso}}$ for Group H and L. The symbols are the same as that in Figure 1.}
   \label{Fig:lightcurve-ADAri}
   \end{center}
\end{figure}

\begin{figure}
   \vspace{2mm}
   \begin{center}
   \hspace{3mm}\psfig{figure=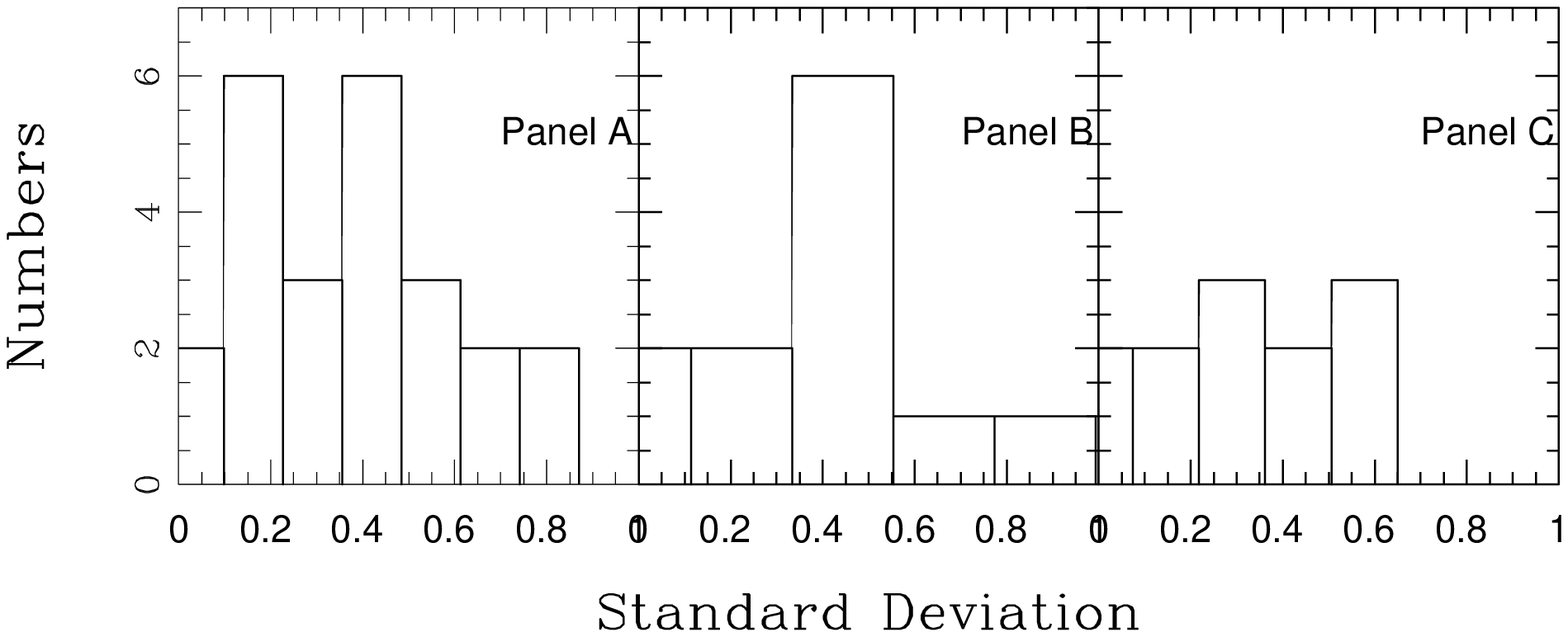,width=100mm,height=100mm,angle=0.0}
   \caption{\it Panel A\rm: Distribution of deviation from the best fit for the whole sample. 
\it Panel B \rm and \it C\rm: The same but for Group H and L, respectively.  
}
   \label{Fig:lightcurve-ADAri}
   \end{center}
\end{figure}

\begin{acknowledgements}
The authors would like to thank the anonymous referee for his/her suggestions and comments 
on the manuscript. We thank Zheng W. K., Prof. Wei J. Y and Hu J. Y. for discussions. 
This work was funded by the NSF of China (NSFC), under grant 10673014.
\end{acknowledgements}


\label{lastpage}


\begin{thebibliography}{99}
\bibitem[Amati 2003]{ama03} Amati L., 2003, ChJAS, 3, 455 
\bibitem[Amati 2006a]{ama06a} Amati L., 2006a, MNRAS, 372, 233
\bibitem[Amati 2006b]{ama06b} Amati L., 2006b, II Nuovo Cimento C, in press (arXiv: astro-ph/0611189v2)
\bibitem[Amati et al. 2002]{ama02} Amati L., Frontera F., Tavani M., et al., 2002, A\&A, 390, 81
\bibitem[Chevalier et al. 2004]{che04} Chevalier R., Li Z., Fransson C., 2004, ApJ, 606, 369
\bibitem[Frail et al. 2001]{fra01} Frail D. A., Kulkarni S. R., Sari R., et al., 2001, ApJL, 562, 55
\bibitem[Fryer et la. 2007]{fry07} Fryer C. L., Mazzali P. A, Prochaska J., et al., 2007, arXiv: astro-ph/0702338
\bibitem[Fynbo et al.(2006)]{fyn06} Fynbo J. P. U., Starling R. L. C., Ledoux C., et al., 2006, A\&A, 451, L47
\bibitem[Ghirlanda et al. 2004]{ghi04} Ghirlanda G., Ghisellini G., Lazzati D., 2004, ApJ, 616, 331
\bibitem[Ghirlanda et al. 2005]{ghi05} Ghirlanda G., Ghisellini G., Firmani C., 2005, MNRAS, 361, L10
\bibitem[Ghirlanda et al. 2007]{ghi07} Ghirlanda G., Nava L., Ghisellini G., et al., 2007, A\&A, 466, 127
\bibitem[Gorosabel et al. 2005]{gor05} Gorosabel J., Jel\'{i}nek M., de Ugarte Postigo A., et al., 2005, NCimC, 28, 677
\bibitem[Hammer et al. 2006]{ham06} Hammer F., Flores H., Schaerer D., et al., 2006, A\&A, 454, 103
\bibitem[Izotov et al. (2006)]{izo06} Izotov Y. I., Papaderos P., Guseva N. G., et al., 2006, A\&A, 454, 137
\bibitem[Kewley et al. 2007]{kew07} Kewley L. J., Brown W. R., Geller M. J., et al., 2007, AJ, 133, 882
\bibitem[Kobulnicky \& Kewley(2004)]{kob04} Kobulnicky H. A., Kewley L. J., 2004, ApJ, 617, 240
\bibitem[Lamb et al. 2004]{lam04} Lamb D. Q., Donaghy T. Q., Graziani C., et al., 2004, NewAR, 48, 459
\bibitem[Le Floc'h et al.(2003)]{lef03} Le Floc'h E., Duc P.-A., Mirabel I. F., et al. 2003, A\&A, 400, 499
\bibitem[Li 2007]{li07}Li L. X., 2007, MNRAS, 379, L55
\bibitem[Liang et al. 2004]{Li04}Liang E. W., Dai Z. G., Wu X. F., 2004, ApJL, 606, 29
\bibitem[Liang et al. 2006]{lih06} Liang Y. C., Hammer F., Flores, H., 2006, A\&A, 447, 113
\bibitem[MacFadyen \& Woosley(1999)]{mac99} MacFadyen A. I., Woosley S. E., 1999, ApJ, 524, 262
\bibitem[Panaitescu 2005]{pan05} Panaitescu A., 2005, MNRAS, 363, 1409
\bibitem[Panaitescu \& Kumar 2002]{pak02} Panaitescu A., Kumar P., 2001, ApJ, 554, 667
\bibitem[Pellizza et al. 2006]{pel06} Pellizza L. J., Duc P. -A., Le Floc'h E., et al., 2006, A\&A, 459, 5
\bibitem[Rau et al. 2005]{rau05} Rau A., Salvato M., Greiner J., 2005, A\&A, 444, 425 
\bibitem[Rees \& M\'{e}sz\'{a}ros 2005]{ree05} Rees M., \& M\'{e}sz\'{a}ros P., 2005, ApJ, 628, 847
\bibitem[Ryde 2005]{ryd05} Ryde F., 2005, ApJL, 625, 95
\bibitem[Sakamoto et al. 2004]{sak04} Sakamoto T., Lamb D. Q., Graziani C., et al., 2004, ApJ, 602, 875
\bibitem[Sari 1999]{sar99} Sari R., 1999, ApJL, 524, 43
\bibitem[Savaglio et al. 2005]{sav05} Savaglio S., Glazebrook K., Le Borgne D., et al., 2005, ApJ, 535, 260 
\bibitem[Savaglio et al.(2006)]{sav06} Savaglio S., Glazebrook K., Le Borgne D., 2006
, in Gamma-Ray Bursts in the Swift Era,
 AIP Conf. Proc., Vol. 838., ed. S. S. Holt, N. Gehrels, \& J. A. Nousek (Melville: American Inst. of Phys.), 540 (arXiv: astro-ph/0601528)
\bibitem[Schaefer 2003]{sch03} Schaefer B. E., 2003, ApJL, 583, 67
\bibitem[Schaefer 2007]{sch07} Schaefer B. E., 2007, ApJ, 660, 16
\bibitem[Soderberg et al. 2007]{sod07} Soderberg A. M., Nakar E., Cenko S. B., et al., 2007, ApJ, 661, 982
\bibitem[Sollerman et al. 2005]{sol05} Sollerman J., \"{O}stlin G., Fynbo J. P. U.,  et al., 2005, NewA, 11, 103
\bibitem[Stanek et al.(2006)]{sta06} Stanek K. Z., Gnedin O. Y., Beacom J. F., et al., 2006, AcA, 56, 333
\bibitem[Thompson 2006]{tho06} Thompson C., 2006, ApJ, 651, 333
\bibitem[Thompson et al. 2007]{thm07} Thompson C., M\'{e}sz\'{a}ros P., Rees M., 2007, ApJ, 666, 1012
\bibitem[Tremonti et al. 2004]{tre04} Tremonti C. A., Heckman T. M., Kauffmann G., et al., 2004, ApJ, 613, 898
\bibitem[Wijers 2001]{wij01} Wijers R., 2001, ``Gamma-Ray Burst in the Afterglow Era'', eds, E. 
Costa, F. Frontera, J. Jorth (Berlin: Springer-Verlag), 306
\bibitem[Woosley 1993]{woo93} Woosley S. E., 1993, ApJ, 405, 273
\bibitem[Woosley \& Bloom(2006)]{woo06a} Woosley S. E., Bloom J. S., 2006, ARA\&A, 44, 507
\bibitem[Zhang \& M\'{e}sz\'{a}ros 2002]{zh02} Zhang B., \& M\'{e}sz\'{a}ros P., 2002, ApJ, 581, 1236



\end{thebibliography}
\end{document}